\documentclass[a4paper,11pt]{article}
\usepackage{pos}
\usepackage{subfig}
\usepackage{graphicx}
\usepackage{lineno}

\title{Anisotropic flow of (multi-)strange hadrons in Au+Au collisions in BES-II energies at STAR.}
\ShortTitle{Anisotropic flow in BES-II}

\author*[a]{Prabhupada Dixit (for the STAR collaboration)}
\affiliation[a]{Indian Institute of Science Education and Research (IISER) Berhampur,\\
  Berhampur, India}

\emailAdd{prabhupadad@iiserbpr.ac.in}
\abstract{In these proceedings, we report the elliptic ($v_{2}$) and triangular ($v_{3}$) flow measurements of strange and multi-strange hadrons such as $K_{S}^{0}$, ~$\Lambda$($\bar{\Lambda}$), ~$\phi$, ~$\Xi^{-}$($\bar{\Xi}^{+}$), and  $\Omega^{-}$($\bar{\Omega}^{+}$) at mid-rapidity ($|y| < 1.0$) in Au+Au collisions at $\sqrt{s_{NN}}$ = 19.6 GeV using high statistics Beam Energy Scan Phase-II (BES-II) data. The transverse momentum ($p_{T}$) dependence, centrality dependence, and the number of constituent quarks (NCQ) scaling of $v_{2}$ and $v_{3}$ are studied for all these particles. A better NCQ scaling is observed in the case of antiparticles compared to particles, which can be attributed to the contribution from the transported quarks in particles. The hydrodynamically motivated ratio ($v_{3}/v_{2}^{3/2}$) is presented as a function of $p_{T}$.
}
\FullConference{%
  41st International Conference on High Energy physics - ICHEP2022\\
  6-13 July, 2022\\
  Bologna, Italy
}

\begin{document}
\maketitle
%\linenumbers

\section{Introduction}
Relativistic heavy-ion collisions provide the opportunity to study matter under extreme conditions of temperature and density where the quarks and gluons are no longer bound inside the hadrons. The azimuthal anisotropic flow of the final-state particles is an important observable for studying the initial stages and hydrodynamic evolution of the medium formed in such collisions. The anisotropic flow can be measured using Fourier series expansion of the azimuthal distributions of the particle yield in the momentum space given by
\begin{equation}
\label{eq-1}
E\frac{d^{3}N}{dp^{3}} =\frac{1}{2\pi}\frac{d^{2}N}{p_{T}dp_{T}dy}\left[1 + \sum_{n} 2v_{n}\cos n(\phi-\psi_{n})\right].
\end{equation}
Here, $v_{n}$ represents the $n^{th}$ order flow coefficient and $\psi_{n}$ is the  $n^{th}$ order event plane angle. The second order coefficient ($v_{2}$) and third order coefficient ($v_{3}$)  are known as elliptic and triangular flow, respectively. The $n^{th}$ order flow coefficient can be measured by
\begin{equation}
\label{eq-2}
v_{n} = \frac{\langle \cos n(\phi -\psi_{n}^{obs})\rangle}{\langle \cos n(\psi_{n}^{obs}-\psi_{n})\rangle}.
\end{equation}
%Here, $\psi_{R}$ is the $n^{th}$ order event plane angle which is a proxy for the true reaction plane angle ($\Psi_{R}$) and the denominator represents the event plane resolution. 
Here, the denominator represents the event plane resolution, the factor that takes the deviation of the observed $n^{th}$ order event plane angle and the true $n^{th}$ harmonic plane angle ($\psi_{n}$).
The angular bracket represents the average over all the particles in the event and over all the events.  Various model studies~\cite{Ref1, Ref2, Ref3} proposed that $v_{2}$ and $v_{3}$ are sensitive to the equation of the state and quantities characterizing the transport properties of the medium, such as the shear viscosity to entropy density ratio ($\eta/s$). 
%Therefore, simultaneous measurements of $v_{2}$ and $v_{3}$ are necessary to constrain the models.
\par
Multi-strange hadrons and $\phi$ mesons have small hadronic interaction cross sections and they freeze out early from the medium~\cite{Ref4}. They are least affected by the late hadronic stage of the medium and are considered as excellent probes for the initial stages of the system. 
%In the late hadronic stage of the medium, the magnitudes of $v_{2}$ and $v_{3}$ can be affected due to hadronic rescattering. To overcome this problem, multi-strange hadrons and $\phi$ mesons are excellent candidates because of their small scattering cross section and early freeze-out from the medium~\cite{Ref4}. 
Prior to the RHIC BES-II program, the STAR detectors were upgraded to have a extended pseudorapidity coverage and better particle identification capabilities. The pseudorapidity coverage becomes wider ($|\eta|$ < 1.5) compared to BES-I ($|\eta|$ < 1.0) by upgrading the inner TPC (iTPC). The statistics of the event sample are improved by a factor of 30 for Au+Au collisions at $\sqrt{s_{NN}}$ =  19.6 GeV compared to BES-I, which enables us to obtain more precise measurements of $v_{2}$ and $v_{3}$ of multi-strange hadrons and $\phi$ mesons at low energy regimes. 

\section{Analysis details}
\begin{figure}[h]
\centering
\includegraphics[width=7cm,clip]{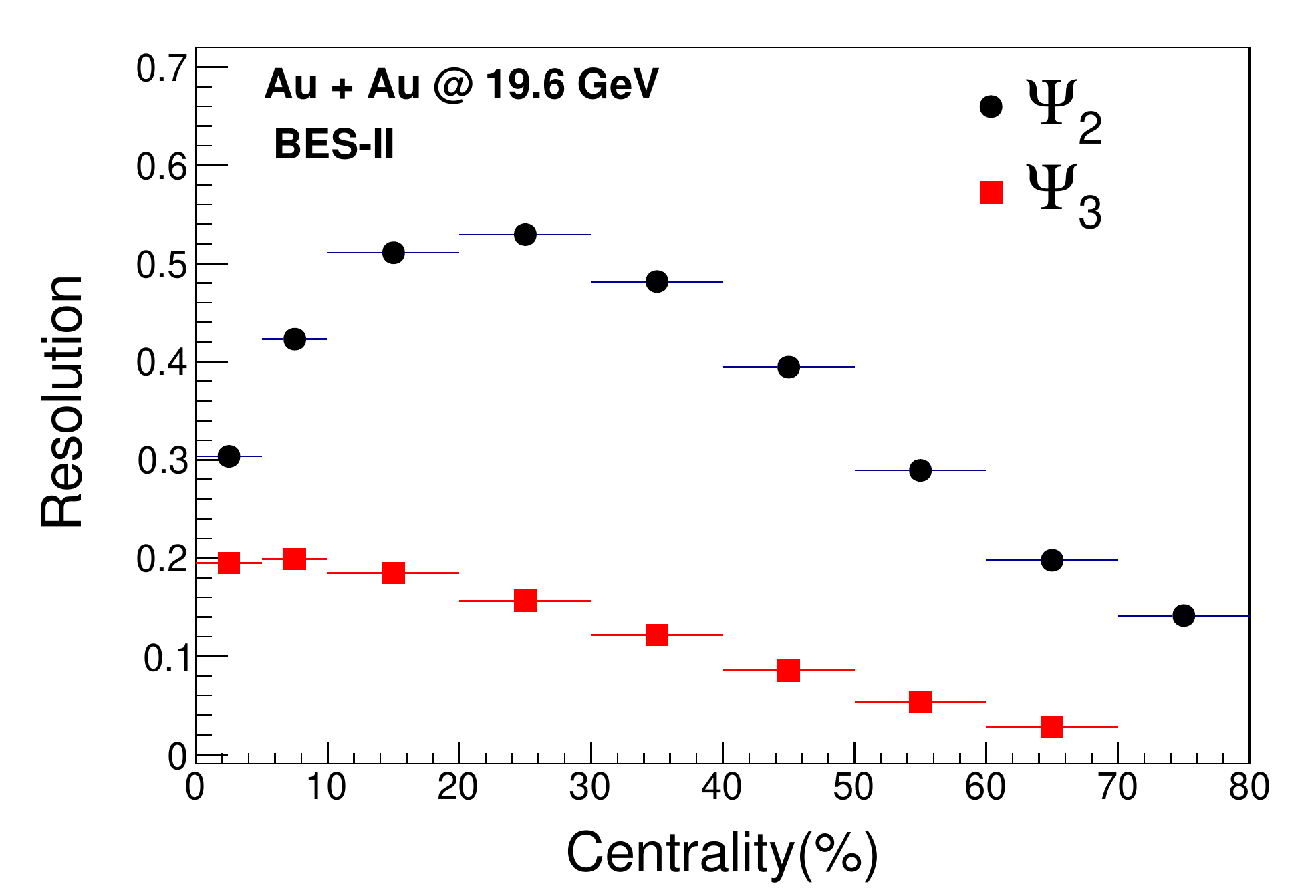}
\caption{Sub-event plane resolutions for $\psi_{2}$ and $\psi_{3}$ are shown as a function of centrality. Both TPC and iTPC are used to construct the event planes.}
\label{fig-1}       % Give a unique label
\end{figure}

First, we need to measure the event plane angle $\psi_{n}$ which is given by
\begin{equation}
\label{eq-3}
\psi_{n} = \frac{1}{n}\tan^{-1}\frac{\sum_{i}w_{i}\sin(n\phi_{i})}{\sum_{i}w_{i}\cos(n\phi_{i})}.
\end{equation}
Here, $\phi_{i}$ is the azimuthal angle of the $i^{th}$ particle. The weight factor $w_{i} = p_{T} \times \phi\text{-}\mathrm{weight}$ is applied to optimize the event plane resolution and to take care of the non-uniform acceptance in $\phi$. Here, the factor for $\phi\text{-}\mathrm{weight}$ is obtained from the inverse of the azimuthal distribution ($dN/d\phi$) of the particles which is used to make the event plane angle distribution isotropic. The details of $\phi\text{-}\mathrm{weight}$ estimation can be found in Ref.~\cite{Ref5}. Experimentally, the event plane resolution is given by
%The event plane is not exactly equal to the true reaction plane; the factor that takes this deviation into account is called the resolution given by
\begin{equation}
\label{eq-4}
R_{n} = \sqrt{\langle\cos n(\psi_{n}^{A} - \psi_{n}^{B})\rangle}.
\end{equation}
Here $\psi_{n}^{A}$ and $\psi_{n}^{B}$ are two sub-event planes in the opposite pseudorapidty ($\eta$) sides of $\eta$ coverage -1.5 < $\eta$ < -0.05 and 0.05 < $\eta$ < 1.5 respectively and $R_{n}$ is the sub-event plane resolution. Figure~\ref{fig-1} shows $R_{n}$ as a function of centrality. The resolution of $\psi_{2}$ is about 10\% higher compared to that of the BES-I measurement~\cite{RefEP}.
%Since the particles we used for the analysis are short-lived, their azimuthal angle cannot be measured directly, therefore, 
We use the invariant mass method~\cite{Ref6} to measure the $v_{n}$ of the short-lived particles used in this analysis. A detailed procedure for the method can be found in Refs.~\cite{Ref7,Ref8}.
\section{Results}
\subsection{$p_{T}$ dependence of $v_{2}$}
\begin{figure}[h]
\centering
\includegraphics[width=11cm,clip]{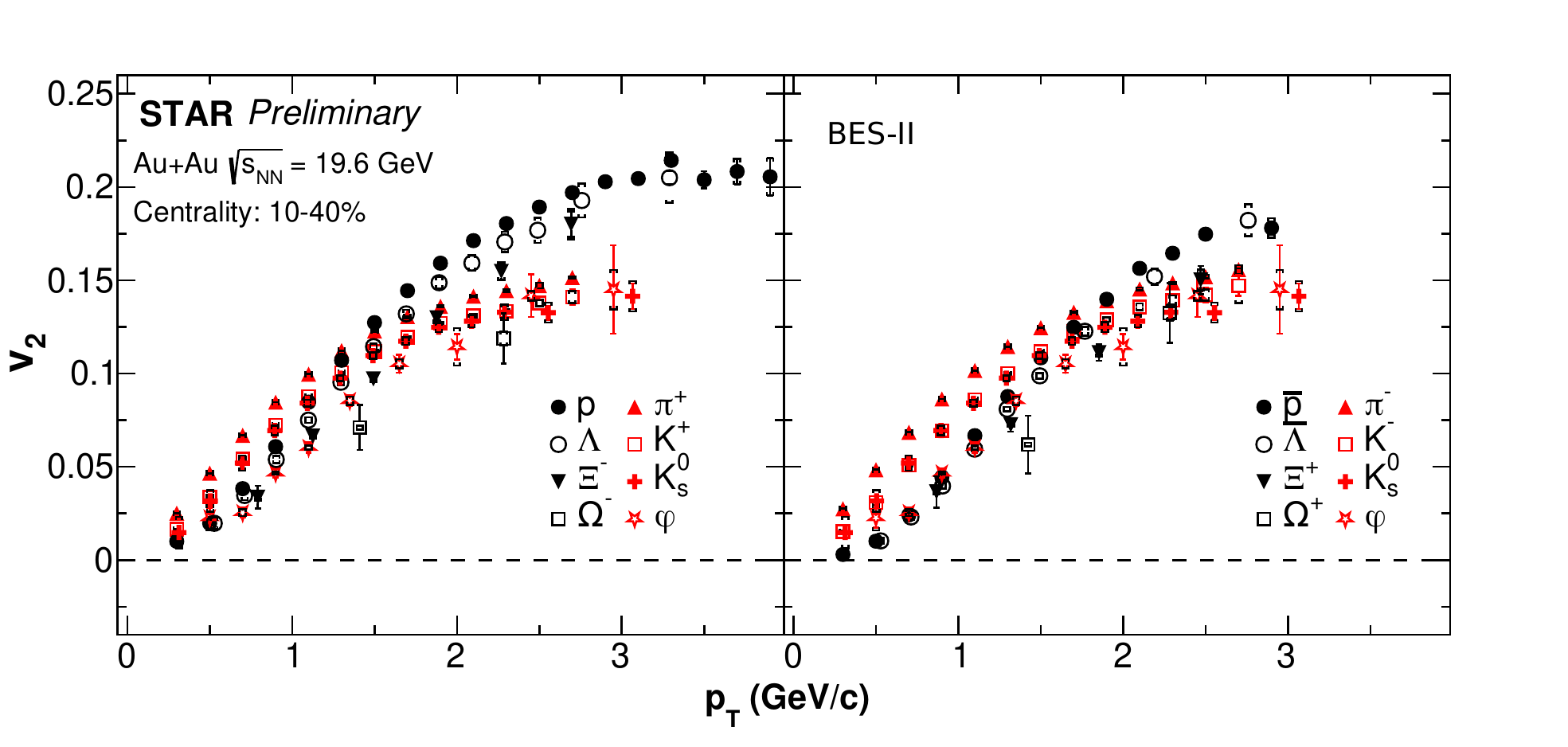}
\caption{The left panel shows $v_{2}$ of particles as a function of $p_{T}$ for the 10-40\% centrality. The right panel shows the same for antiparticles. The vertical lines and caps represent the statistical uncertainties and systematic uncertainties, respectively.}
\label{fig-2}       % Give a unique label
\end{figure}

Figure~\ref{fig-2} shows $v_{2}$ of particles and anti-particles as a function of $p_{T}$ in 10-40\% centrality for all the (multi-)strange hadrons such as $K^{+}(K^{-})$, $K_{S}^{0}$, $\Lambda$($\bar{\Lambda}$), $\Xi^{-}$($\bar{\Xi}^{+}$), $\Omega^{-}$($\bar{\Omega}^{+}$), and $\phi$ mesons along with non-strange hadrons such as $\pi^{+}(\pi^{-})$ and $p(\bar{p})$. We observe a mass ordering in the lower $p_{T}$ ($p_{T}$ < 1.5 GeV/$c$) region, that is, lighter mass particles show higher $v_{2}$ due to the radial flow of the system~\cite{Rad_flow}. Above $p_{T}$ > 1.5 GeV/$c$, baryon-meson separation is observed, consistent with the picture of the coalescence model of hadronization~\cite{Coal_mod}. The statistical error bars in the BES-II measurements are about three times smaller than those in the BES-I measurements.
\subsection{Centrality dependence of $v_{2}$ and $v_{3}$}
\begin{figure}[h]
\centering
\includegraphics[width=11cm,clip]{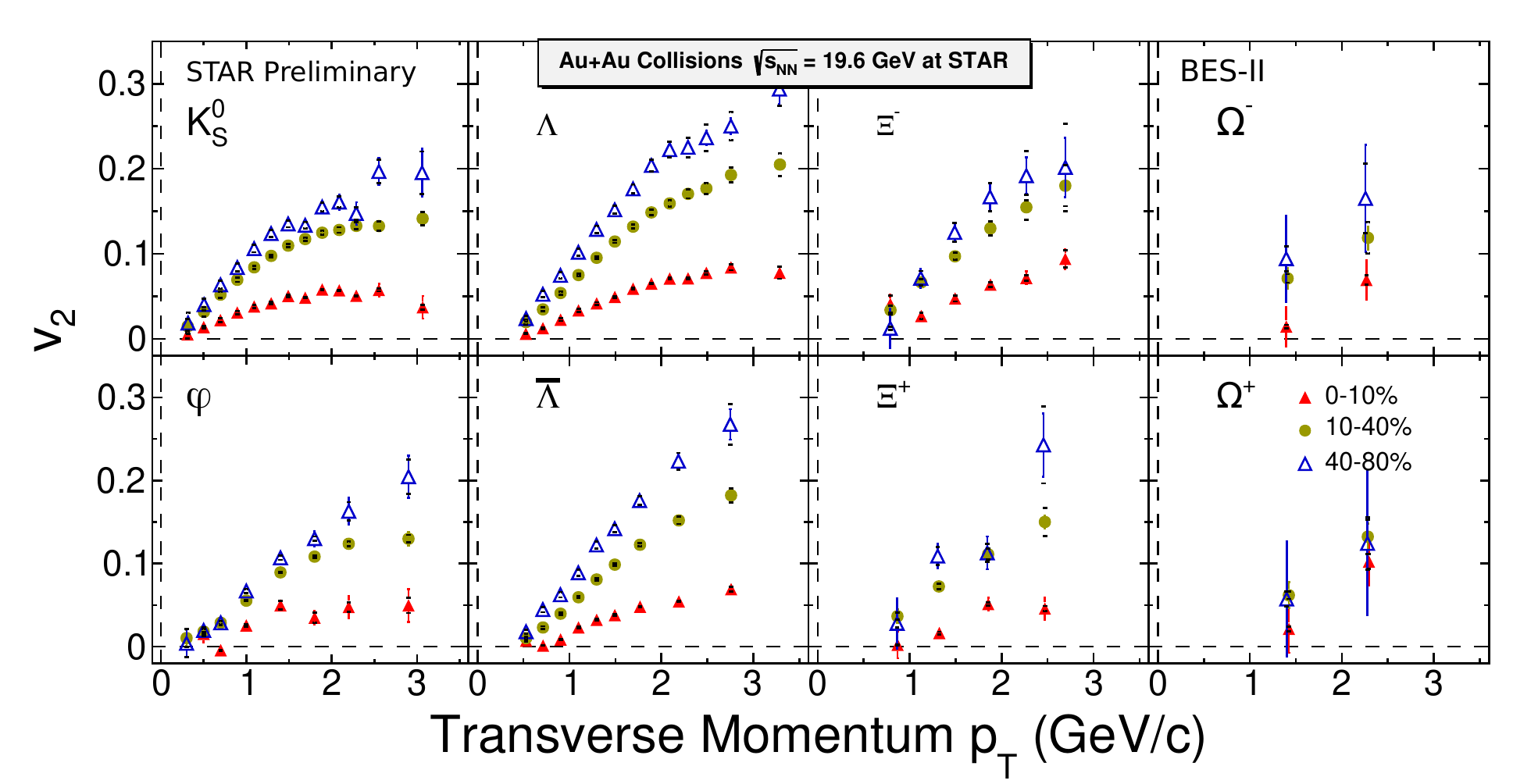}
\caption{$v_{2}$ as a function of $p_{T}$ for three different centrality classes, 0-10\%, 10-40\%, and 40-80\%. The vertical lines and caps represent the statistical uncertainties and systematic uncertainties, respectively.}
\label{fig-3}       % Give a unique label
\end{figure}

\begin{figure}[h]
\centering
\includegraphics[width=11cm,clip]{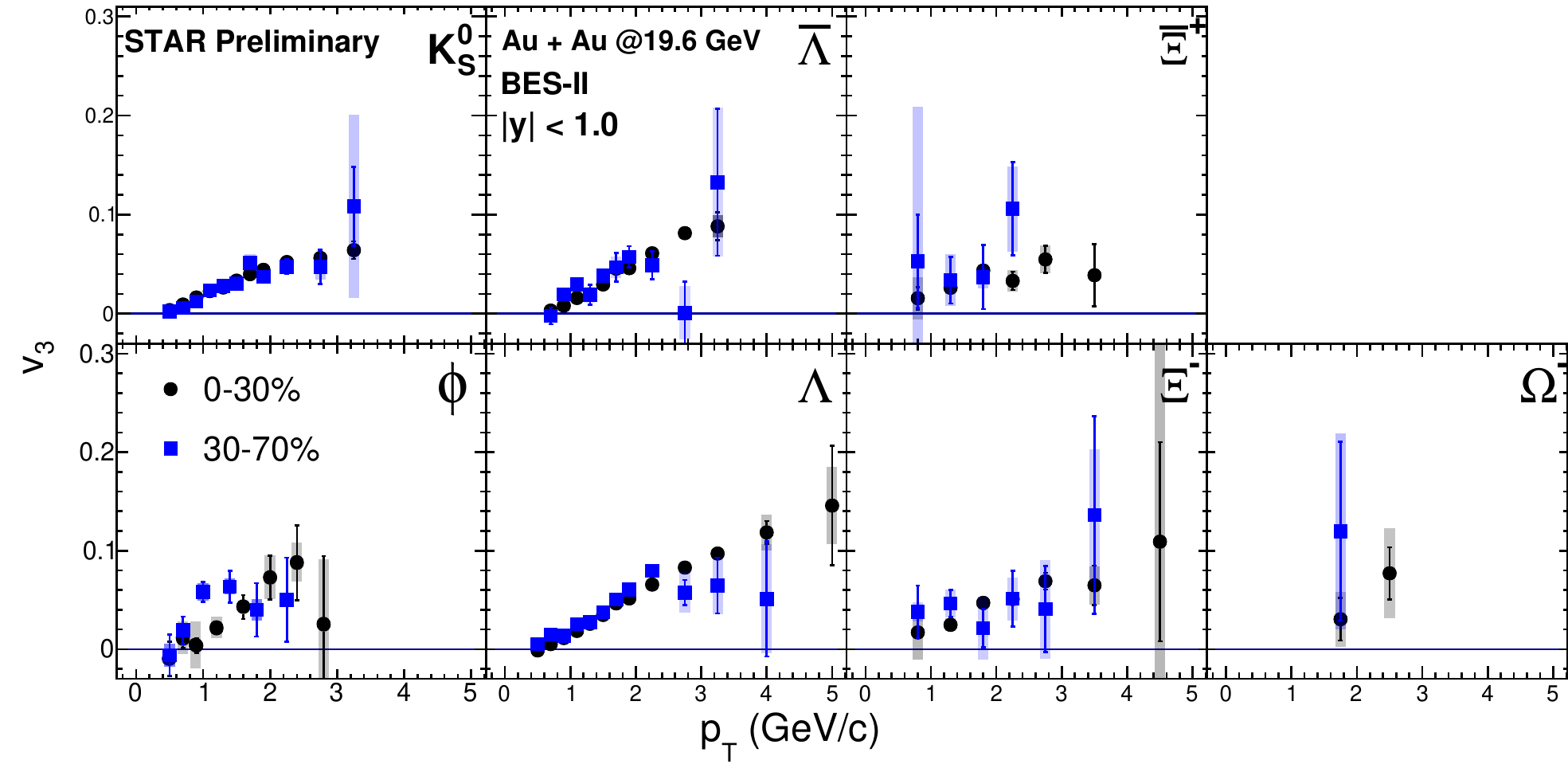}
\caption{$v_{3}$ as a function of $p_{T}$ for two different centrality classes, 0-30\% and 30-70\%. The vertical lines and shaded boxes represent the statistical uncertainties and systematic uncertainties, respectively.}
\label{fig-4}       % Give a unique label
\end{figure}

The centrality dependence of $v_{2}$ and $v_{3}$ are studied. As shown in Fig.~\ref{fig-3}, $v_{2}$ shows a strong centrality dependence for all particles.
%$v_{2}$ shows strong centrality dependence for all particles, as shown in Fig.~\ref{fig-3}. 
This indicates that $v_{2}$ arises predominantly from spatial anisotropy in the overlap region of the colliding nuclei.
The centrality dependence of $v_{3}$ is weak, as shown in Fig.~\ref{fig-4}. This is due to the fact that the dominant cause of $v_{3}$ is the event-by-event fluctuation of the energy density in the overlap region of the two nuclei.

\subsection{Number of constituent quarks scaling in $v_{3}$}
\begin{figure}[h]
\centering
\subfloat{{\includegraphics[width=7cm]{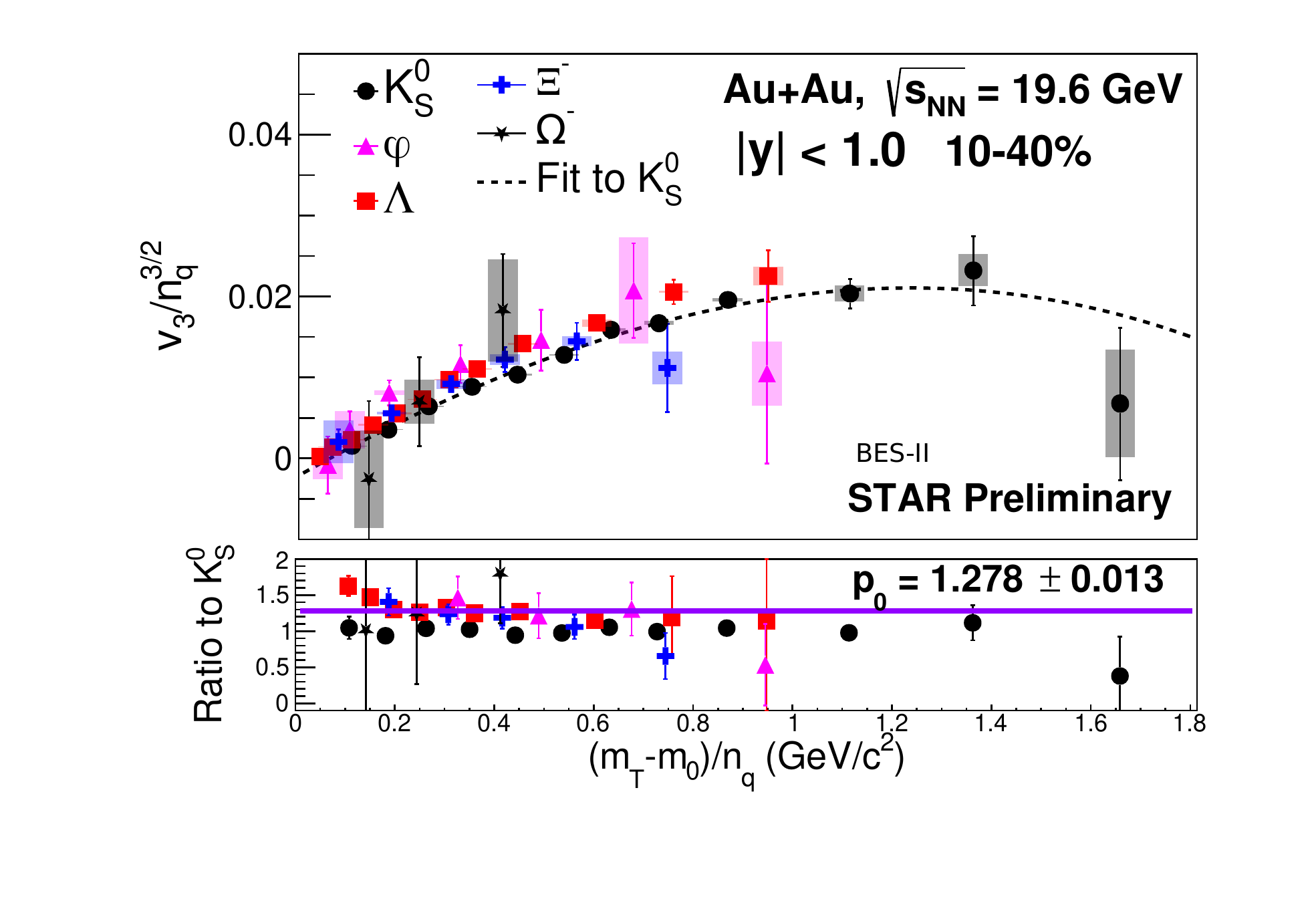} }}
%\qquad
\subfloat{{\includegraphics[width=7cm]{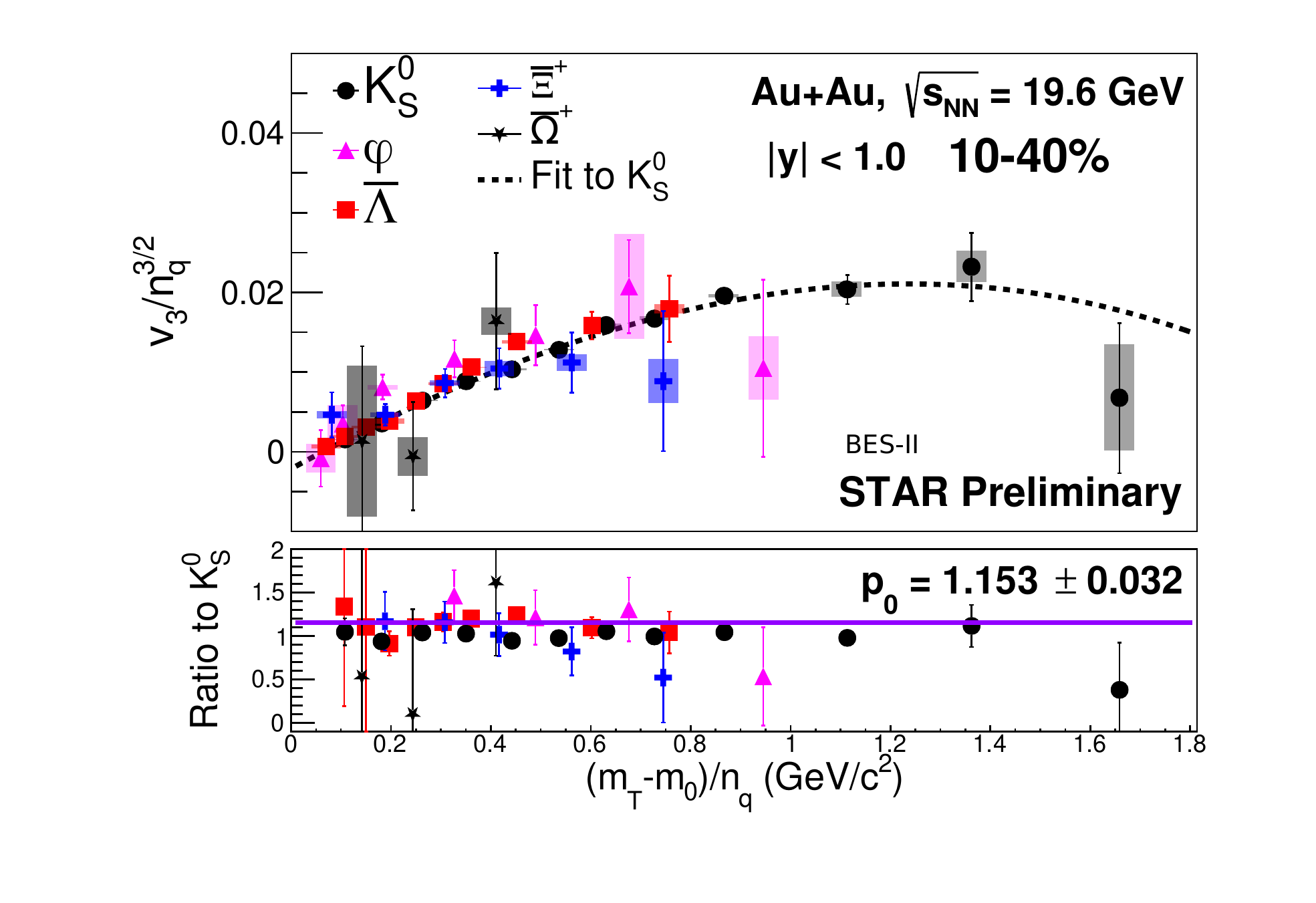} }}
\caption{ Left figure shows the $v_{3}/n_{q}^{3/2}$ as a function of the NCQ scaled $m_{T}-m_{0}$ for particles in 10-40\% centrality class. The lower left panel shows the straight line fit to the scaled $v_{3}$ ratios taken with respect to the $K_{S}^{0}$ fit line. Right figure shows the same for the antiparticles. The vertical lines and shaded boxes represent the  statistical uncertainties and systematic uncertainties, respectively.}
\label{fig-5}       % Give a unique label
\end{figure}
Figure~\ref{fig-5} shows the modified NCQ-scaled $v_{3}$, i.e., $v_{3}/n_{q}^{3/2}$~\cite{ncq_v3}, as a function of NCQ-scaled transverse kinetic energy $(m_{T} - m_{0})/n_{q}$ for all particles (left) and antiparticles (right), respectively. The $K_{S}^{0}$ data points are fitted with a third-order polynomial and the ratios of $v_{3}/n_{q}^{3/2}$ of all other particles (antiparticles) are taken with respect to the fit function of $K_{S}^{0}$. To quantify the scaling, the ratios are fitted simultaneously with a polynomial of order zero. It is observed that the scaling holds within 15\% for antiparticles and within 30\% for particles. The scaling is better for antiparticles than for particles; this could be due to the effect from transported quarks in particles. The picture is consistent with NCQ-scaled $v_{2}$ measured from BES-II data at 19.6 GeV~\cite{Ref9}. 
%The NCQ-scaling is also measured at 14.6 GeV using the minimum bias events with event statistics ~400M. The scaling holds within 20\% for $v_{2}$. Measurements in finer centrality bins at this energy are ongoing.

\subsection{\textbf{$v_{3}/v_{2}^{3/2}$} ratio}
\begin{figure}[h]
\centering
\includegraphics[width=12cm,clip]{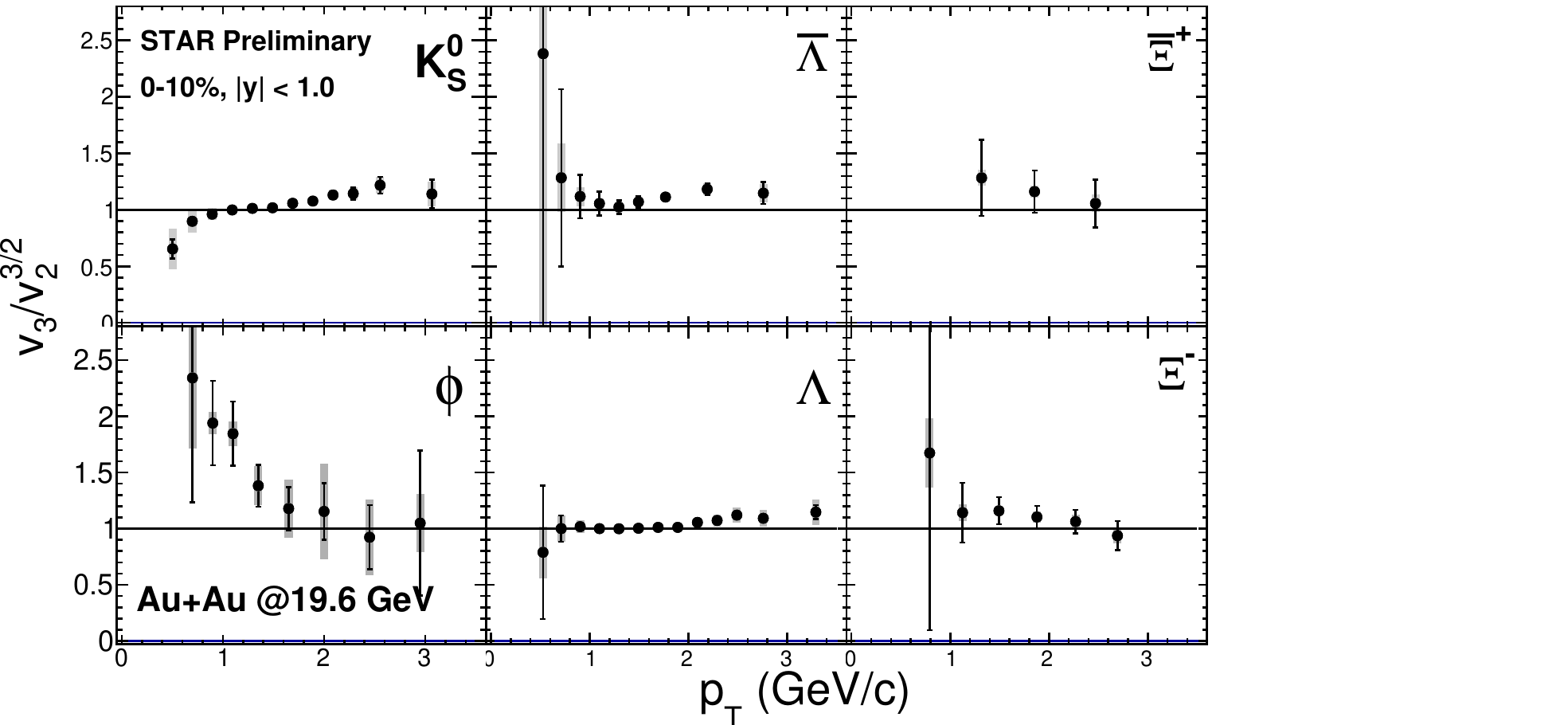}
\caption{$v_{3}/v_{2}^{3/2}$ as a function of $p_{T}$ in the 10-40\% centrality for $K_{S}^{0}$, $\Lambda$($\bar{\Lambda}$), $\phi$, and  $\Xi^{-}$($\bar{\Xi}^{+}$). The vertical lines and shaded boxes represent the  statistical uncertainties and systematic uncertainties, respectively.}
\label{fig-rat}       % Give a unique label
\end{figure}

The hydrodynamics-motivated ratio of $v_{3}/v_{2}^{3/2}$ is suggested to be independent of $p_{T}$ and its magnitude is sensitive to the transport properties of the medium~\cite{Ref10, Ref11, Ref12}. Figure~\ref{fig-rat} shows the ratio $v_{3}/v_{2}^{3/2}$ as a function of $p_{T}$ in 10-40\% centrality. The ratio shows a weak dependence on $p_{T}$ above 1.5 GeV/$c$ for all the particle species.

\section{Summary}
In summary, we report the elliptic flow and triangular flow of (multi-)strange hadrons and $\phi$ meson in Au+Au collisions at $\sqrt{s_{NN}}$ = 19.6 GeV using the high-statistics BES-II data. The $p_{T}$ dependence of $v_{2}$ shows a mass ordering at $p_{T}$ < 1.5 GeV/$c$ and a particle-type separation at $p_{T}$ > 1.5 GeV/$c$, which depends on the valence quark content of the particle species. The centrality dependence of $v_{2}$ and $v_{3}$ is studied. The $v_{2}$ shows strong centrality dependence for all the particle species to be driven by the centrality dependence of initial ellipticity. Unlike $v_{2}$, the centrality dependence in $v_{3}$ is weak because it is expected to driven be by initial state fluctuations. The modified NCQ scaling for $v_{3}$ is found to hold better for antiparticles compared to particles, which can be attributed to the effect of transported quarks initially present in the colliding system. A weak $p_{T}$ dependence of hydrodynamics-motivated ratio,  $v_{3}/v_{2}^{3/2}$, is observed for all the particles above $p_{T}$ > 1.5 GeV/$c$.

\end{document}